\documentclass[prl,twocolumn,superscriptaddress]{revtex4}
\usepackage{amsmath}
\usepackage{amsfonts}
\usepackage{mathrsfs}
\usepackage{graphicx}
\usepackage{times}
\usepackage{color}

\begin{document}

\title{Weyl Superfluidity in a Three-dimensional Dipolar Fermi Gas}
\author{Bo Liu}
\email{liubophy@gmail.com}
\affiliation{Wilczek Quantum Center,
Zhejiang University of Technology, Hangzhou 310023, China}
\affiliation{Department of Physics and Astronomy, University of
Pittsburgh, Pittsburgh, PA 15260, USA}
\author{Xiaopeng Li}
\affiliation{Condensed Matter Theory Center and Joint Quantum Institute, University of
Maryland, College Park, MD 20742, USA}
\author{Lan Yin}
\affiliation{School of Physics, Peking University, Beijing 100871, China}
\author{W. Vincent Liu}
\email{w.vincent.liu@gmail.com}
\affiliation{Wilczek Quantum Center,
Zhejiang University of Technology, Hangzhou 310023,
China}
\affiliation{Department of Physics and Astronomy, University
of Pittsburgh, Pittsburgh, PA 15260, USA}

\begin{abstract}
Weyl superconductivity or superfluidity, a fascinating topological
state of matter, features novel phenomena such as emergent Weyl
fermionic excitations and anomalies. Here we report that an
anisotropic Weyl superfluid state can arise as a low temperature
stable phase in a 3D dipolar Fermi gas. {A crucial ingredient of our
model is a direction-dependent two-body effective attraction
generated by a rotating external field.} Experimental signatures are
predicted for cold gases in radio-frequency spectroscopy. The finite
temperature phase diagram of this system is studied and the
transition temperature of the Weyl superfluidity is found to be
within the experimental scope for atomic dipolar Fermi gases.
\end{abstract}

\maketitle

Weyl superfluids or semimetals represent recent developments in
generalizing topological phases from gapped to gapless systems
(e.g., from topological insulators to semimetals), in condensed matter physics~\cite%
{2003_Volovik_book,2011_Leon_Physics}. These Weyl states are characterized
by the presence of two (or more) gapless Weyl points, which are
topologically protected against small perturbations. The Weyl nodes lead to
a variety of fascinating phenomena such as unusual surface states~\cite%
{2011_wan_PRB,2014_Fan_PRL}, Hall
effects~\cite{2011_Yang_PRB,2011_Xugang_PhysRevLett}, and other
transport features~\cite{2011_Leon_PRL,2012_Ashvin_PRL}. {Finding
electronic materials supporting Weyl states has attracted
considerable interests~\cite{2014_Ashvin_Review}. There are many
proposed potential candidate materials, such as the pyrochlore iridates~\cite%
{2011_wan_PRB}, topological insulator multilayer structures~\cite%
{2011_Leon_PRL,2012_Leon_PhysRevB,2012_Burkov_PhysRevB,2012_Meng_PhysRevB},
as well as certain quasicrystals~\cite{2013_Timusik_PhysRevB}.
However, there is still no compelling experimental evidence for the
observation of one.} In the field of ultracold atoms, this phase was
predicted to appear in spin-orbit coupled Fermi
gases~\cite{2011_Gongming_PRL,2014_xuyong_PRL}. This line of active
research awaits for the future experimental breakthrough of
synthesizing higher dimensional artificial spin-orbit coupling with
controlled heating~\cite{2013_Spielman_nature}. After all, the
search for Weyl superconductors remains an open problem for both
electronic and ultracold atomic systems.

In this letter, we report the emergence of Weyl superfluidity in a 3D
single-component dipolar Fermi gas with an effective attraction engineered
by a rotating external field. Recently, degenerate dipolar Fermi gases
witnessed rapid developments in both magnetic dipolar atoms (such as $^{167}$%
Er ~\cite{2014_Grimm_PRL,2014_Ferlaino_arxiv} and $^{161}$Dy~\cite%
{2012_Luming_PRL,2014_Benjamin_PRA} atoms) and polar molecules~\cite%
{2008_Junye_Science,2012_Martin_PRL}, stimulating tremendous
interests in dipolar effects in many-body phases. The effects of the
anisotropic dipolar interaction on the fermion many-body physics
have been extensively investigated~\cite{2012_Baranov_Reviews}. {In
particular, this provides the possibility of superfluid pairing
between dipolar Fermi atoms in spinless or multicomponent
systems~\cite{2012_congjunwu_Scirep,2010_congjunwu_PhysRevB,2013_Huizhai_PRL,2013_SuYi_PhysRevLett}
at low temperatures.} For dipoles aligned parallel to the $z$
direction, a $p$-wave superfluid state with the dominant $p_z$
symmetry was studied in a three-dimensional dipolar Fermi
gas~\cite{1999_Liyou_PRA} and the competition between this
superfluidity and nematic charge-density-wave (CDW) was also
discussed~\cite{2014_Lanyin_PhysRevB}. For a dipolar Fermi gas
confined in a 2D plane, superfluid states of $p$-wave symmetry~\cite%
{2008_Bruun_PRL,2012_Boliu_PRA,2009_Cooper_PRL}, including a $p+ip$ state in
particular~\cite{2012_Boliu_PRA,2009_Cooper_PRL}, are predicted.

{The key idea here is to engineer a direction-dependent two-body
effective attraction, which supports Cooper pairs with the chirality
encoded in the $p$-wave pairing gap. This Weyl superfluid state
breaks time reversal symmetry as well as inversion symmetry.} Such
broken symmetries have profound implications for the interesting
topological defects~\cite{2003_Volovik_book}. {We shall describe
this state in a 3D magnetic dipolar Fermi gas composed of one
hyperfine sate, which has been realized in the experimental system
of $^{167} $Er~\cite{2014_Grimm_PRL} recently. The direction of
dipole moments can be fixed by applying an external magnetic field.
Let the external field be orientated at a small angle with respect
to the $xy$-plane and rotate fast around the $z$ axis. {The
time-averaged interaction between dipoles~\cite{2002_Pfau_PRL} is
isotropically attractive in the $xy$-plane and repulsive in the
$z$-direction. In general, the attraction is expected to cause
Cooper pairing instability while the repulsion should restrict the
pairing from certain nodal directions. Their combined effect could
give rise to Weyl Fermi points for the Bogoliubov quasi-particles.
Such a heuristically argued result is indeed confirmed by a
self-consistent calculation through the model to be introduced
below.}

\textit{Effective model.} Consider a 3D spinless dipolar Fermi gas
subjected to an external rotating magnetic field
\begin{equation*}
\textstyle
\mathbf{B}(t)=B[{\hat{z}}\cos {\varphi }+\sin {\varphi
}({\hat{x}}\cos {\Omega }t+{\hat{y}}\sin {\Omega }t)],
\end{equation*}%
where $\Omega $ is the rotation frequency, $B$ is the magnitude of magnetic
field, the rotation axis is $z$, and $\varphi $ is the angle between the
magnetic field and the $z$ axis. In strong magnetic fields, dipoles are
aligned parallel to $\mathbf{B}(t)$. With fast rotations, the effective
interaction between dipoles is the time-averaged interaction
\begin{equation*}
\textstyle
V(\mathbf{r})={\frac{d^{2}(3\cos ^{2}\varphi
-1)}{2r^{3}}(1-3\cos ^{2}\theta )}\equiv \frac{d^{\prime
2}}{r^{3}}{(1-3\cos ^{2}\theta )}, \label{Interaction}
\end{equation*}
where $d^{\prime 2}\equiv d^{2}{\frac{(3\cos ^{2}\varphi -1)}{2}}$
with the magnetic dipole moment $d$, {$\mathbf{r}$ is the vector
connecting two dipolar particles, and $\theta$ is the angle between $\mathbf{%
r}$ and the $z$ axis.} The effective attraction, $V(\mathbf{r})<0$, is
created by making $\cos \varphi <\sqrt{1/3}$, which is our focus in this
work.

The effective Hamiltonian of the system above is given by $H=\int
d^{3}\mathbf{r\psi }^{\dagger }(\mathbf{r})[-\frac{\hbar
^{2}\bigtriangledown ^{2}}{2m}-\mu ]\mathbf{\psi }(\mathbf{r})
+\frac{1}{2} \int d^{3}\mathbf{r}\int d^{3}\mathbf{r}^{\prime }\mathbf{%
\psi }^{\dagger }(\mathbf{r})\mathbf{\psi }^{\dagger }(\mathbf{r}^{\prime
})V(\mathbf{r-r}^{\prime })\mathbf{\psi }(\mathbf{r}^{\prime })\mathbf{\psi }%
(\mathbf{r})$, where $\psi(\mathbf{r})$ is the fermion field and
$\mu$ is the chemical potential.

{Due to the attractive interaction, fermions tend to pair with each
other and form a superfluid state at low temperatures. To study this
superfluid state, we construct a general theory to describe a
spinless Fermi gas by a fully self-consistent
Hartree-Fock-Bogoliubov method. The details are given in
Supplementary Materials. Constructing a bosonic effective action by
Hubbard-Stratonovich transformation, we obtain self-consistent
equations under a saddle-point approximation for the fermion
bilinears $\kappa (\mathbf{r}) =\int d^{3}\mathbf{r}^{\prime}V(
\mathbf{r-r}^{\prime})  \psi ^{\dag}(\mathbf{r}^{\prime}) \psi
(\mathbf{r}^{\prime})$, $\lambda (\mathbf{r,r}^{\prime })
=-V(\mathbf{r-r}^{\prime}) \psi ^{\dag}(\mathbf{r}) \psi
(\mathbf{r}^{\prime})$, and
$\tilde{\Delta}(\mathbf{r,r}^{\prime }) =V(\mathbf{r-r}%
^{\prime}) \psi (\mathbf{r}^{\prime }) \psi (\mathbf{r})$.
Correspondingly, the Hartree-Fock self-energy and superconducting
gap are given  as
\begin{eqnarray}
\textstyle
\Sigma (\mathbf{r}^{\prime }\mathbf{,r}) &\equiv &\langle \kappa (\mathbf{r}%
)\rangle \delta (\mathbf{r} - \mathbf{r}^{\prime }) +\langle \lambda (%
\mathbf{r}^{\prime }\mathbf{,r})\rangle ,  \notag \\
\Delta (\mathbf{r}^{\prime }\mathbf{,r}) &\equiv &\langle \tilde{\Delta}(%
\mathbf{r}^{\prime }\mathbf{,r})\rangle ,  \label{SFGap}
\end{eqnarray}
where $\langle \ldots \rangle$ means the expectation value in the ground state.
}

\textit{3D uniform dipolar Fermi gas.} We now apply the general
theory outlined above to the system of a 3D uniform spinless dipolar
Fermi gas in the presence of a rotating magnetic field. From the
symmetry of the system, at least for not too strong interaction
strength, we anticipate that pairing only occurs between a particle
with momentum $\mathbf{k}$ and another with momentum $-\mathbf{k}$
as in the standard BCS theory. Due to the translational symmetry, it
is convenient to study this problem in the momentum space. After
Fourier transformation of Eq.~\eqref{SFGap}, the Hartree-Fock
self-energy and the pairing gap read
\begin{equation}
\textstyle
\Sigma _{\mathbf{k}}=V(0)n-{\frac{1}{\upsilon }}\sum_{\mathbf{k}^{\prime }}V(%
{\mathbf{k}-\mathbf{k}^{\prime }}){\frac{{1}}{{2}}}\left[1-\frac{\xi _{%
\mathbf{k}^{\prime }}}{E_{\mathbf{k}^{\prime }}}\tanh ({\frac{\beta }{2}}E_{%
\mathbf{k}^{\prime }})\right],  \label{SFenergy1}
\end{equation}
\begin{equation}
\textstyle
\Delta _{\mathbf{k}}=-{\frac{1}{\upsilon }}\sum_{\mathbf{k}^{\prime }}V({%
\mathbf{k}-\mathbf{k}^{\prime }})\frac{\Delta _{\mathbf{k}^{\prime }}}{2E_{%
\mathbf{k}^{\prime }}}\tanh ({\frac{\beta }{2}}E_{\mathbf{k}^{\prime }}),
\label{GAP1}
\end{equation}
where $E_{\mathbf{k}}$ is the quasi-particle excitation energy given by $E_{%
\mathbf{k}}=\sqrt{\xi _{\mathbf{k}}^{2}+|\Delta _{\mathbf{k}}|^{2}}$ with
the kinetic energy of fermions $\xi _{\mathbf{k}}=\varepsilon _{\mathbf{k}%
}+\Sigma _{\mathbf{k}}-\mu $ and $\varepsilon _{\mathbf{k}}={\frac{\hbar
^{2}k^{2}}{2m}}$. The interaction between two dipoles in the momentum space
is given by $V(\mathbf{q})={\frac{4\pi d^{\prime 2}}{3}}(3\cos ^{2}\theta _{%
\mathbf{q}}-1)$, with the angle $\theta _{\mathbf{q}}$ between momentum $%
\mathbf{q}$ and $z$ axis, $n$ is the total density, $\upsilon $ is the
volume, and $\beta =1/(k_{B}T)$.

It is known that the gap equation (Eq.~\eqref{GAP1}) has ultraviolet
divergence~\cite{2013_Huizhai_PRL}. The origin of the divergence can be
attributed to the singularity of the dipolar interaction potential for large
momentum, or equivalently for short distance. Just as in the treatment of
two-component Fermi gas with contact interaction~\cite{2008_Stringari_RMP},
we need to regularize the interaction in the gap equation (Eq.~\eqref{GAP1}%
). The divergence can be eliminated by expressing the bare interaction $V(%
\mathbf{k}-\mathbf{k}^{\prime })$ in Eq.~\eqref{GAP1} in terms of the vertex
function (scattering off-shell amplitude)~\cite{2002_Baranov_PRA} as $\Gamma (\mathbf{k}
-\mathbf{k}^{\prime })=V(\mathbf{k}-\mathbf{k}^{\prime })-{%
\frac{1}{\upsilon }}\sum_{\mathbf{q}}\,\Gamma (\mathbf{k}-\mathbf{q})\frac{1%
}{2\varepsilon _{q}}V(\mathbf{q}-\mathbf{k}^{\prime })$, and the gap
equation will be renormalized as
\begin{equation}
\textstyle \Delta (\mathbf{k})=-{\frac{1}{\upsilon }} \sum
_{\mathbf{k}^{\prime }} \Gamma (\mathbf{k}-\mathbf{k}^{\prime
})\Delta (\mathbf{k}^{\prime
})\left[ \frac{\tanh \frac{\beta E(\mathbf{k}^{\prime })}{2}}{2E(\mathbf{k}%
^{\prime })}-\frac{1}{2\varepsilon _{k^{\prime }}}\right] \,.  \label{GAP2}
\end{equation}%
Note that the Hartree term for the selfenergy in Eq.~\eqref{SFenergy1}, $%
V(0)n $ vanishes, since for dipolar interaction in 3D uniform system, $%
V(0)=0 $~\cite{2010_Congjun_PhysRevA} and renormalization of the interaction
has a negligible effect on the self-energy. Then, the Hartree-Fock
self-energy is expressed as
\begin{equation}
\textstyle
\Sigma _{\mathbf{k}}=-{\frac{1}{\upsilon }}\sum_{\mathbf{k}^{\prime }}V({{%
\mathbf{k}-\mathbf{k}^{\prime }}}){\frac{{1}}{{2}}} \left[ 1-\frac{\xi _{%
\mathbf{k}^{\prime }}}{E_{\mathbf{k}^{\prime }}}\tanh ({\frac{\beta }{2}}E_{%
\mathbf{k}^{\prime }})\right].  \label{SFenergy2}
\end{equation}

The total density $n$ can be obtained from the thermodynamic potential $%
\Omega$ by using the relation $N=-\partial \Omega/\partial {\mu}$,
\begin{equation}
\textstyle
n=\sum_{\mathbf{k}}{\frac{{1}}{{2\upsilon}}}\left[1-\frac{\xi_{\mathbf{k}}}{%
E_{\mathbf{k}}}\tanh({\frac{\beta }{2}}E_{\mathbf{k}})\right].
\label{Number}
\end{equation}

Under the constraint of Fermi statistics for this single component
dipolar Fermi gas, the dominant pairing instability is in the
channel with orbital angular momentum $L= 1$. The most stable low
temperature phase has $p_x+ip_y$ symmetry, following from the fact
that this phase fully gaps the Fermi surface, in contrast to
competing phases, such as $p_x$ or $p_y$ superfluid
state~\cite{1961_Anderson_PhysRev}. Note that in the presence of a
rotating magnetic field, all the dipoles rotating with respect to
the $z$-axis, so the system has a $SO(2)$ spatial rotation symmetry.
This symmetry is not broken in the $p_x + ip_y$ pairing state, and
we can thus write down the
Cooper pair as $\Delta_{\mathbf{k}}\equiv\Delta(k_\rho,k_z)e^{i\varphi_{%
\mathbf{k}}}$, where $k_\rho=\sqrt{k_x^2+k_y^2}$ and
${\varphi_{\mathbf{k}}}$ is the polar angle of the momentum
$\mathbf{k}$ in the $xy$-plane, to simplify the calculation in
Hartree-Fock-Boguliubov approach.

\textit{Weyl fermions.} With the time-reversal symmetry
spontaneously broken in the superfluid state, topological properties
emerge in
quasi-particle excitations, which are described by a mean field Hamiltonian $%
H_{SF}=\sum_{\mathbf{k}}[\xi _{\mathbf{k}}c_{\mathbf{k}}^{\dagger }c_{%
\mathbf{k}}+{\frac{\Delta _{\mathbf{k}}^{\ast }}{2}}c_{-\mathbf{k}}c_{%
\mathbf{k}}+{\frac{\Delta _{\mathbf{k}}}{2}}c_{\mathbf{k}}^{\dagger }c_{-%
\mathbf{k}}^{\dagger }]$, with $c_{\mathbf{k}}$ the fermion
annihilation operator. {This Hamiltonian can be expressed in the
matrix form by
\begin{eqnarray*}
\textstyle
H_{SF} &=&\sum_{\mathbf{k}}(c_{\mathbf{k}}^{\dag },c_{-\mathbf{k}})%
\begin{pmatrix}
{\frac{{\xi (\mathbf{k})}}{2}} & {\frac{{\Delta (\mathbf{k})}}{2}} \\
{\frac{{{\Delta ^{\ast }}(\mathbf{k})}}{2}} & {\ -\frac{{\xi (\mathbf{k})}}{2%
}}%
\end{pmatrix}%
\begin{pmatrix}
c_{\mathbf{k}} \\
c_{-\mathbf{k}}^{\dag }%
\end{pmatrix}
\\
&\equiv &\sum_{\mathbf{k}}(c_{\mathbf{k}}^{\dag },c_{-\mathbf{k}})\vec{d}(%
\mathbf{k})\cdot \vec{\sigma}\,%
\begin{pmatrix}
c_{\mathbf{k}} \\
c_{-\mathbf{k}}^{\dag }%
\end{pmatrix}.
\end{eqnarray*}%
where the $\vec{d}$ vector is defined in terms of the Pauli matrices
$\sigma $'s.} {The $d_{x,y}$ components vanish along the $k_z$ axis,
whereas along this axis, $d_{z}$ vanishes at only two points
${\mathbf k}_{+}^{C}=(0,0,k_{+}^{C})$ and ${\mathbf
k}_{-}^{C}=(0,0,k_{-}^{C})$($=-{\mathbf k}_{+}^{C}$)
(Fig.~\ref{figure3}a).} {In the $(k_{x},k_{y})$ momentum plane with
$k_{-}^{C}<k_{z}<k_{+}^{C}$, $\vec{d}({\bf k})$ wraps around a
sphere as shown in Fig.~\ref{figure3}b. {Evidently, it points} to
the south pole on the $k_z$ axis, while with increasing $k_\rho$,
the $\vec{d}({\bf k})$ vector varies continuously and eventually
points to the north pole as $k_\rho \to \infty$. This vector
$\vec{d}({\bf k})$  thus forms a skyrmion in the momentum space with
a topological charge $\pm 1$ (where the `$\pm $' sign reveals the
spontaneous time-reversal symmetry breaking). However, in other
regions $k_{z}>k_{+}^{C}$ or $k_{z}<k_{-}^{C}$, the topological
charge vanishes.} These two gapless points ${\bf k}_{\pm }^{C}$ are
Weyl nodes, defining the corresponding topological transitions in the momentum space~\cite%
{2011_Leon_PRL,2014_xuyong_PRL}. Close to the Weyl nodes, the
Hamiltonian takes the form of $2\times 2$ Hamiltonian of a chiral
Weyl fermion~\cite{1929_Weyl_ZF}. We have checked that the
quasi-particle energy dispersion $E_{\mathbf{k}}$ is linear around
both two Weyl points, for instance as shown in Fig.~\ref{figure2}b
when the interaction strength $J=3$
where $J\equiv |\frac{md^{\prime 2}}{\hbar ^{2}}k_{F}|$. As shown in Fig~\ref%
{figure2}c and d, the Weyl nodes are hedgehog-like topological
defects of the vector field $\vec{d}(\mathbf{k})$, which are the
point source of Berry flux in momentum space, with a
topological invariant $N_{C}=\pm 1$. Here $N_{C}$ is defined by $N_{C}=\frac{%
1}{24\pi ^{2}}\epsilon _{\mu \nu \gamma \chi }$tr$\oint\nolimits_{\Sigma }d%
\mathbb{S}^{\chi }G\frac{\partial G^{-1}}{\partial k_{\mu }}G\frac{\partial
G^{-1}}{\partial k_{\nu }}G\frac{\partial G^{-1}}{\partial k_{\gamma }}$,
where $G^{-1}$ is the inverse Green's function for the quasi-particle excitation, $%
\Sigma $ is a 3D surface around the isolated Fermi point
$\mathbf{k}_{+}^{C}$ or $\mathbf{k}_{-}^{C}$, and \mbox{tr} stands
for the trace over the relevant particle-hole degrees of
freedom~\cite{2003_Volovik_book}. The quasi-particle excitations
near the Fermi points realize the long-sought low-temperature analog
of Weyl fermions as originally proposed in particle physics. These
Weyl nodes are separated from each other in momentum space. They can
not be hybridized, which makes them indestructible, as they can only
disappear by mutual annihilation of pairs with opposite topological
charges. This is the mechanism of topological stability of this Weyl
superfluid state, which is distinct from the spectral-gap protection
in insulating topological phases. {To characterize the existence of
Weyl fermions, we calculate the fermionic density of states (DOS)
for superconducting
states~\cite{2012_Muller_PRA,1997_Morita_PhysRevLett}
$N(E)=\frac{1}{{{{(2\pi )}^{3}}}}\int {{d^{3}}\mathbf{k}}\frac{1}{2}(1+\frac{{\xi (\mathbf{k}%
)}}{{E(\mathbf{k})}})\delta (E-E(\mathbf{k}))$, which is directly
related to the radio frequency (rf) spectroscopy
signal~\cite{2011_Han_PhysRevA}. With linear dispersion near Weyl
nodes, we find  $N(E)\propto E^{2}$ when $E\rightarrow 0$, which is
a direct manifestation of Weyl fermions. This behavior of DOS is
confirmed in our numerics (Fig.~\ref{figure2}a). The experimental
advances in rf measurement~\cite{2003_Jin_PRL,2003_Ketterle_Science}
makes the detection of this signal experimentally accessible.}

The other important feature of Weyl fermions realized in this
dipolar gas is that they have anisotropic dispersion, reflecting the
anisotropy of dipolar interactions. In Fig.~\ref{figure2}b, {the
conic quasi-particle dispersion as a function of the momentum ${{\bf
k}-{\mathbf k}^C_+}$ is shown. This momentum is chosen with a
certain angle $\tilde{\theta}$ respecting to the $k_z$ axis. The
cones with positive and negative branches  correspond to the
Bogoliubov quasi-particle energy $\pm E({\bf k}-{\mathbf k}^C_+)$.
The Fermi velocity, shown by the slope of the quasi-particle
dispersion, strongly depends on the angle $\tilde{\theta}$.}

\begin{figure}[t]
\begin{center}
\includegraphics[scale=0.15]{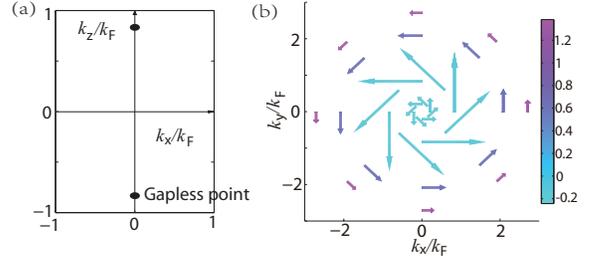}
\end{center}
\caption{(a) Gapless points along the $k_z$ axis, where the unit of
momentum is the Fermi momentum $k_F$. (b) {Illustration of the
skyrmion configuration formed by $\vec{d}({\bf k})$ vector in the
$(k_x,k_y)$ plane, with fixed $k_z \in (k_- ^C, k_+ ^C)$. The arrows
show $d_{x,y}$ components, and the colors index the $d_z$
component.}} \label{figure3}
\end{figure}

\begin{figure}[t]
\begin{center}
\includegraphics[scale=0.2]{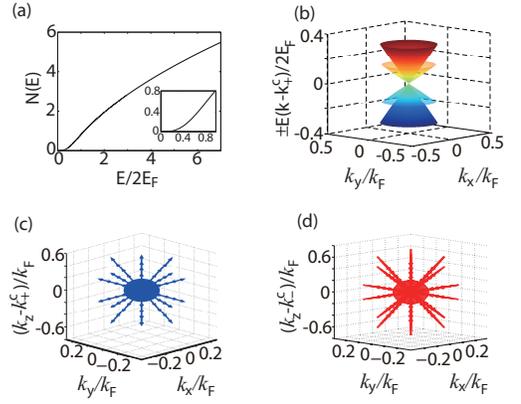}
\end{center}
\caption{(a) Density of states (DOS) which has been defined in the
main text in units of $n_F/E_F$, where $n_F=\frac{k_F^3}{6\pi^2}$
and $E_F=\frac{\hbar^2k_F^2}{2m}$. (b) {Quasi-particle dispersion
around the gapless points. There are four branches of conic energy
spectra shown here. For the two branches in the middle we choose
$\tilde {\theta}  = \pi/10$, while for the other two we choose
$\pi/2$. (c) and (d) Hedgehog-like topological defects formed by the
$\vec{d} ({\bf k})$ vector around two Weyl nodes.}} \label{figure2}
\end{figure}

\textit{Anisotropic superconducting gap.} We now discuss the
superconducting gap for fermions resulting from anisotropic
dipole-dipole interaction. For clarity of demonstration, we take the
first-order Born approximation by replacing the vertex function
$\Gamma(\mathbf{k}-\mathbf{k}^{\prime})$ in
the gap equation (Eq.~\eqref{GAP2}) by the bare dipolar interaction $V(%
\mathbf{k}-\mathbf{k}^{\prime})$. By numerically solving the Hartree-Fock
self-energy equation (Eq.~\eqref{SFenergy2}), the gap equation (Eq.~%
\eqref{GAP2}), and number equation (Eq.~\eqref{Number}) self-consistently,
the superconducting gap anisotropy has been investigated. As shown in Fig.~%
\ref{figure1}a, the magnitude of the order parameter
(superconducting gap) on the Fermi surface $\Delta_{F}(\theta_{\bf
k})$ monotonically increases when enlarging the angle $\theta_{\bf
k}$ between the momentum $\mathbf{k}$ and $z$ axis. The maximum
value of $\Delta_{F}(\theta_{\bf k})$ is reached in the direction
perpendicular to the dipoles, say $\theta_{\bf k}={\frac{\pi }{2}}$.
{This is because the dipolar interaction is mostly attractive when
$\theta_{\bf k}={\frac{\pi }{2}}$.} In the direction of the dipoles,
namely $\theta_{\bf k}=0$ the order parameter vanishes.
{Fig.~\ref{figure1}b shows that the order parameter is also
dependent on $k_{\rho}$ with fixed $k_z$.} This can be understood
from the analysis of the gap equation (Eq.~\eqref{GAP2}) that the
main contribution to the integral comes from the region of small
momentum which is close to the Fermi surface. {In the weak
interaction regime, the pairing order parameter is exponentially
small, for instance when $J=3$ it is around $10^{-3}E_F$. However,
when the interaction strength increases, the superconducting gap
will be comparable to $E_F$. For example when $J=15$ it reaches
around $0.4E_F$.} The anisotropy of the order parameter provides a
crucial difference from both $s$~\cite{2008_Stringari_RMP} and
$p$-wave pairing~\cite{2007_Leo_AnnalsPhy} due to a short-range
attractive interaction. This anisotropy ensures the anisotropic
momentum dependence of the gap in the spectrum of single particle
excitations. For example, excitations with momenta perpendicular to
the direction of the dipoles acquire the largest gap. In contrast to
this, the excitations with momenta in the direction of the dipoles
remain unchanged. Therefore, the response of this dipolar superfluid
Fermi gas to small external perturbations will have a pronounced
anisotropic character.

\begin{figure}[t]
\begin{center}
\includegraphics[width=8cm]{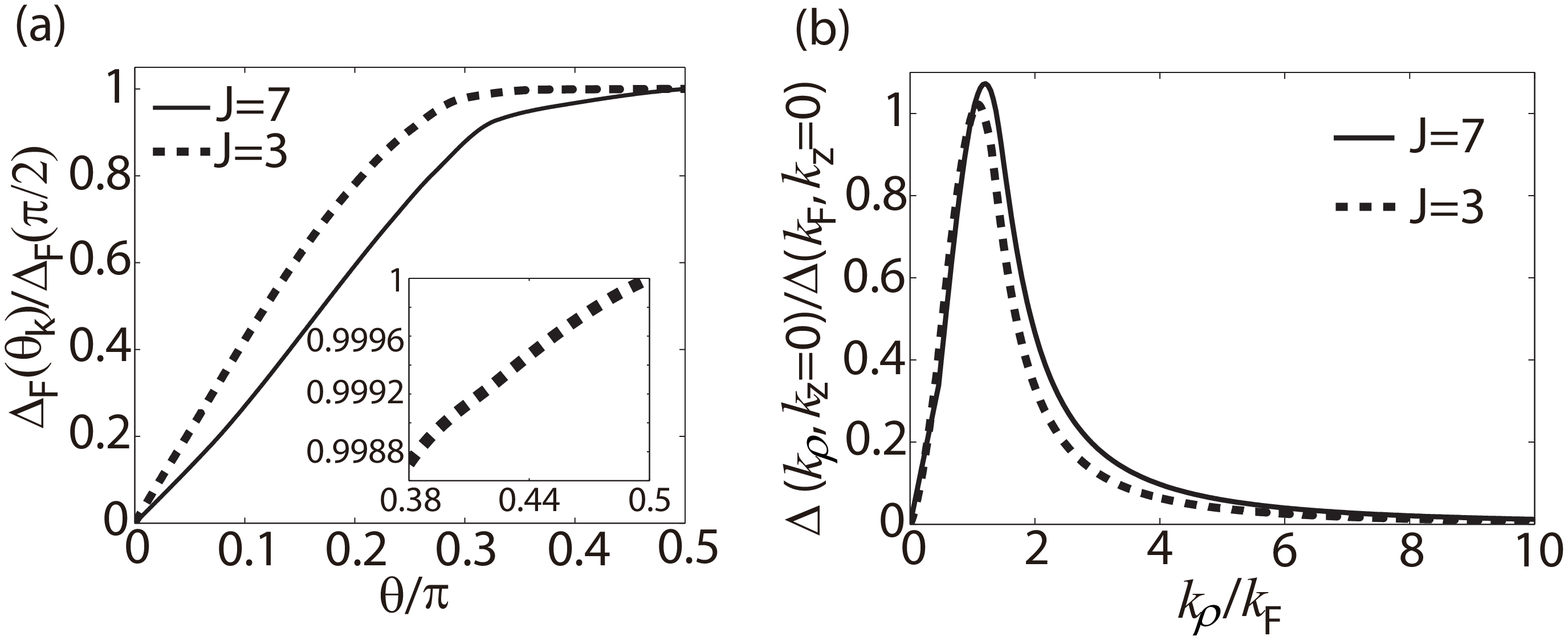}
\end{center}
\caption{{Anisotropic superconducting pairing order parameter with
different interaction strengths $J=3$ and $7$ ($J\equiv
|\frac{md^{\prime 2}}{\hbar ^{2}}k_{F}|$). (a) The superconducting
gap $\Delta_{F}(\theta_{\bf k})$ on the Fermi surface versus the
angle $\theta_{\bf k}$ between the momentum $\mathbf{k}$ and $z$
axis. (b) The superconducting gap $\Delta(k_\rho,k_z)$ as a function
of ${k_\rho}$ with fixed $k_z$.}} \label{figure1}
\end{figure}

\textit{Finite temperature phase transition.} {Upon increasing temperature
the Weyl superfluid state will undergo a phase transition to a normal state.
} By numerically solving the Hartree-Fock self-energy equation (Eq.~%
\eqref{SFenergy2}), gap equation (Eq.~\eqref{GAP2}), and number equation
(Eq.~\eqref{Number}) self-consistently at finite temperature, the BCS
transition temperature is obtained as shown in Fig.~\ref{figure5}. We find
that the BCS transition temperature is a monotonically increasing function
of the interaction strength $J$. However, the strong enough interaction will
cause the system to suffer from the mechanical instability. The reason for
that is as follows. The magnitude of superconducting gap increases with
enhancing the interaction strength. Due to the attractive nature of the
effective interaction between dipoles, the free energy of this dipolar gas
is smaller than that of an ideal Fermi gas. This energy reduction increases
with the interaction strength (or equivalently the density of the gas with a
certain dipole moment). When the interaction strength is large enough, the
effect of the interaction is dominant and the system can be unstable. As
shown in Fig.~\ref{figure4}, the chemical potential is a monotonically
decreasing function when the density is above a critical value, and the
compressibility is negative, indicating that the superfluid state is
dynamically unstable. By considering the mechanical instability of the
system, as shown in Fig.~\ref{figure5}, the finite temperature phase diagram
is obtained. We find that the BCS transition temperature of a stable
superfluid state can reach around $0.2E_F$ at mean-field level, which
approaches to the current experimental temperature region~\cite%
{2014_Grimm_PRL,2012_Luming_PRL}.

\begin{figure}[t]
\begin{center}
\includegraphics[scale=0.2]{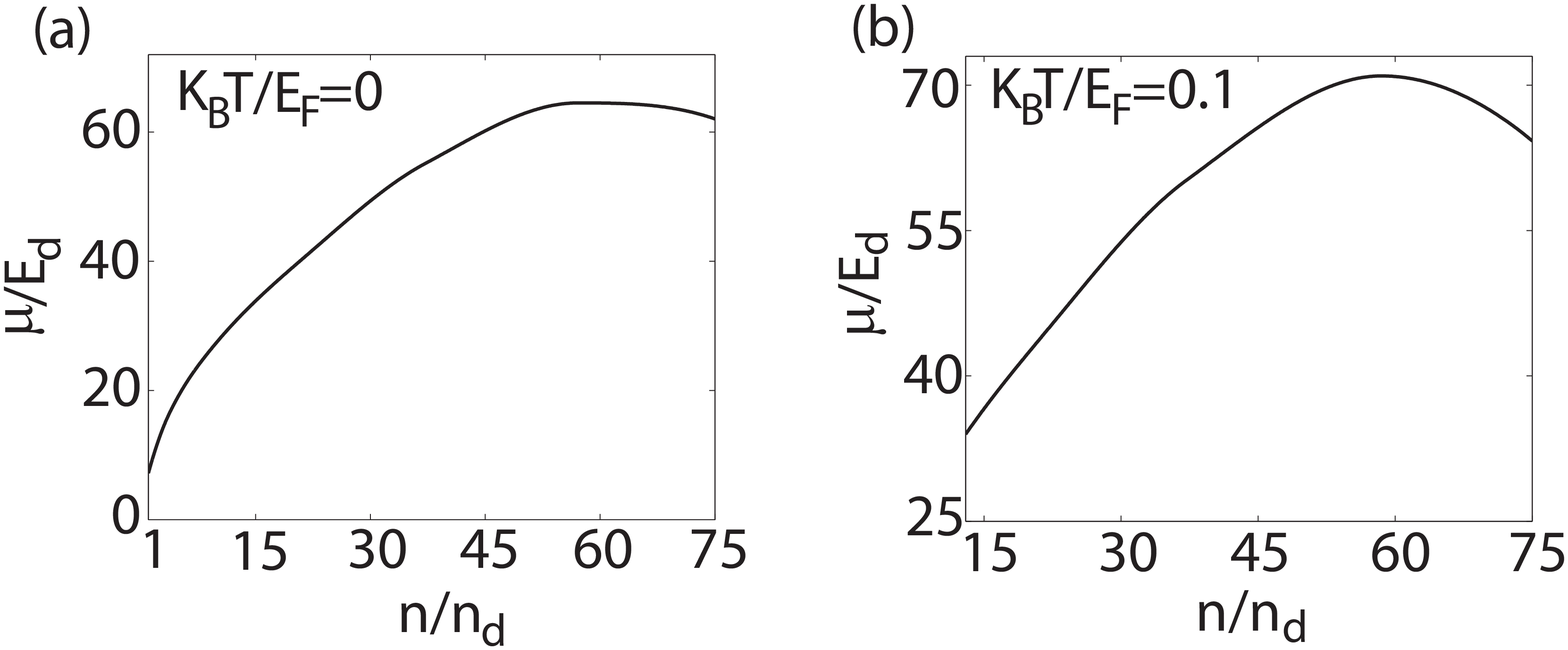}
\end{center}
\caption{Chemical potential $\protect\mu$ versus the density $n$. In (a),
the temperature is $T=0$, while in (b) the temperature is $k_BT=0.1E_F$.
Here, the unit of $\protect\mu$ is $E_d\equiv\hbar^6/(m^3d^4)$ and the unit
of n is $n_d\equiv[\hbar^2/(md^2)]^3$.}
\label{figure4}
\end{figure}

\begin{figure}[t]
\begin{center}
\includegraphics[scale=0.35]{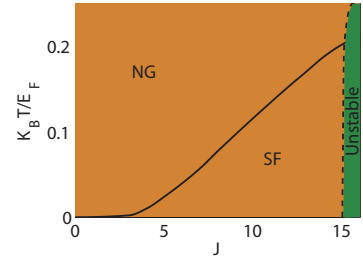}
\end{center}
\caption{Finite temperature phase diagram---The solid line stands for the
BCS transition temperature which separates the region between the superfluid
state (SF) and normal state (NG). The area on the right hand side of the
dash line demonstrates the instability of the system due to the strongly
attractive interaction.}
\label{figure5}
\end{figure}

{In the current experiments, for example, $^{167}$Er atom's magnetic
dipole moment is $7{\mu_B}$ and the density of the
system is about $n=4\times 10^{14} cm^{-3}$. The Fermi energy is given by $%
E_F=\frac{\hbar^2}{2m}(6\pi^2n)^{2/3}\approx 0.16$MHz and the
corresponding Fermi temperature is $T_F=\frac{E_F}{k_B}\approx
1$$\mu$K. To increase the effective attraction, one may consider
adding a shallow optical lattice. For instance with lattice strength
$V=6E_R$, the BCS transition temperature can reach around $3$nK. A
similar estimate can be obtained for $^{161}$Dy atom which has a
larger magnetic dipole moment of $10{\mu_B}$, the corresponding
dipolar interaction strength is around two times larger than that of
$^{167}$Er. Under the same condition, the BCS transition temperature
can reach around $50$nK. {Furthermore, taking advantage of recent
experimental realization of Feshbach resonance in magnetic
lanthanide atoms such as Er~\cite{2014_Frisch_nature}, the
dipole-dipole interaction is highly tunable. The transition
temperature is estimated to reach around 0.2$\mu$K or even higher.}
This high transition temperature $T_c$ makes it promising to obtain
the Weyl superfluid state in experiments.

\textit{Conclusion.} We propose that an anisotropic Weyl superfluid
state can be realized in a 3D spinless dipolar Fermi gas. The
crucial ingredient of our model is the direction-dependent effective
attraction between dipoles generated by a rotating external field.
The long-sought low-temperature analog of Weyl fermions of particle
physics has been found in the quasi-particle excitations in this
superfluid state. The stability and the transition temperature are
also studied, which will be useful for exploring this Weyl
superfluid state in future experiments.}

\textit{Acknowledgements.} This work is supported by AFOSR
(FA9550-12-1-0079), ARO (W911NF-11-1-0230), DARPA OLE Program through ARO,
the Charles E. Kaufman Foundation and The Pittsburgh Foundation (B.L. and
W.V.L.). X.L. acknowledges support by JQI-NSF-PFC and ARO-Atomtronics-MURI.
L.Y. is supported by NSFC under Grant No. (11274022).

\bibliographystyle{apsrev}
\bibliography{wfdipolar}

\begin{widetext}

\begin{center}
{\Large\bf Supplementary Materials}
\end{center}

\renewcommand{\thesection}{S-\arabic{section}} \renewcommand{\theequation}{S%
\arabic{equation}} \setcounter{equation}{0} 
\renewcommand{\thefigure}{S\arabic{figure}} \setcounter{figure}{0}

\section{Path integral approach}
By introducing Grassmann fields $\phi (\mathbf{r,}\tau )$ and $\phi
^{\ast }(\mathbf{r,}\tau )$, which represent fermion fields,
the grand partition function of the system is expressed as (the units are chosen as $%
\hbar =k_{B}=1$)
\begin{equation}
Z=\int D\phi D\phi ^{\ast }e^{-S},  \label{Partition}
\end{equation}%
with the action $$ S[\phi ,\phi ^{\ast }]=S_{0}[\phi ,\phi ^{\ast
}]+S_{int}[\phi ,\phi ^{\ast }],$$ and $$S_{0}[\phi ,\phi ^{\ast
}]=\int d\tau
\int d^{3}\mathbf{r}\int d^{3}\mathbf{r}^{\prime }\phi ^{\ast }(\mathbf{r,}%
\tau )[\frac{\partial }{\partial \tau }-\frac{\bigtriangledown
^{2}}{2m}-\mu ]\delta (\mathbf{r-r}^{\prime })\phi
(\mathbf{r}^{\prime }\mathbf{,}\tau ),$$
$$S_{int}[\phi ,\phi ^{\ast
}]=\frac{1}{2}\int d\tau \int d^{3}\mathbf{r}\int
d^{3}\mathbf{r}^{\prime }\phi ^{\ast }(\mathbf{r,}\tau )\phi ^{\ast }(%
\mathbf{r}^{\prime }\mathbf{,}\tau )V(\mathbf{r-r}^{\prime })\phi (\mathbf{r}%
^{\prime }\mathbf{,}\tau )\phi (\mathbf{r,}\tau ).$$ The quartic
fermionic interaction term in the action in Eq.~\eqref{Partition}
can be decoupled by
introducing Hubbard-Stratonovich fields $\kappa (\mathbf{r,}\tau ),$ $%
\lambda (\mathbf{r,r}^{\prime },\tau ),$ and $\tilde{\Delta}(\mathbf{r,r}%
^{\prime },\tau ).$ This leads to a partition function with the
action
\begin{eqnarray*}
&&S[\phi ,\phi ^{\ast },\kappa ,\lambda ,\lambda ^{\ast },\tilde{\Delta},%
\tilde{\Delta}^{\ast }]=-\int d\tau \int d^{3}\mathbf{r}\int d^{3}\mathbf{r}%
^{\prime }\{\frac{1}{2}[\kappa (\mathbf{r,}\tau ) \\
&&V^{-1}(\mathbf{r-r}^{\prime })\kappa (\mathbf{r}^{\prime },\tau )+\frac{%
|\lambda (\mathbf{r,r}^{\prime },\tau )|^{2}}{V(\mathbf{r-r}^{\prime })}+%
\frac{|\tilde{\Delta}(\mathbf{r,r}^{\prime },\tau )|^{2}}{V(\mathbf{r-r}%
^{\prime })}]\} \\
&&-\int d\tau \int d^{3}\mathbf{r}\int d^{3}\mathbf{r}^{\prime
}[\phi ^{\ast
}(\mathbf{r,}\tau ),\phi (\mathbf{r,}\tau )]\mathbf{G}^{-1}[%
\begin{tabular}{l}
$\phi (\mathbf{r}^{\prime }\mathbf{,}\tau )$ \\
$\phi ^{\ast }(\mathbf{r}^{\prime }\mathbf{,}\tau )$%
\end{tabular}%
\ ],
\end{eqnarray*}

where
\begin{equation*}
\mathbf{G}^{-1}(\mathbf{r,r}^{\prime },\tau )=\frac{1}{2}(\mathbf{G}%
_{0}^{-1}(\mathbf{r,r}^{\prime },\tau )-[%
\begin{tabular}{ll}
$0$ & $\tilde{\Delta}(\mathbf{r,r}^{\prime },\tau )$ \\
$-\tilde{\Delta}^{\ast }(\mathbf{r,r}^{\prime },\tau )$ & $0$%
\end{tabular}%
]),
\end{equation*}

with
\begin{equation*}
\mathbf{G}_{0}^{-1}(\mathbf{r,r}^{\prime },\tau )=[%
\begin{tabular}{ll}
$G_{0}^{-1}(\mathbf{r,r}^{\prime },\tau )$ & $0$ \\
$0$ & $-G_{0}^{-1}(\mathbf{r}^{\prime }\mathbf{,r},\tau )$%
\end{tabular}%
],
\end{equation*}%
and
\begin{equation*}
G_{0}^{-1}(\mathbf{r,r}^{\prime },\tau )=-[(\frac{\partial }{\partial \tau }-%
\frac{\bigtriangledown ^{2}}{2m}-\mu +\kappa (\mathbf{r,}\tau ))\delta (%
\mathbf{r-r}^{\prime })+\lambda (\mathbf{r}^{\prime
}\mathbf{,r},\tau )].
\end{equation*}

After integrating out the fermion fields, the effective action is
obtained

\begin{eqnarray*}
&&S_{eff}[\kappa ,\lambda ,\lambda ^{\ast },\tilde{\Delta},\tilde{\Delta}%
^{\ast }]=-\int d\tau \int d^{3}\mathbf{r}\int d^{3}\mathbf{r}^{\prime }%
\frac{1}{2}[\kappa (\mathbf{r,}\tau ) \\
&&V^{-1}(\mathbf{r-r}^{\prime })\kappa (\mathbf{r}^{\prime },\tau )+\frac{%
|\lambda (\mathbf{r,r}^{\prime },\tau )|^{2}}{V(\mathbf{r-r}^{\prime })}+%
\frac{|\tilde{\Delta}(\mathbf{r,r}^{\prime },\tau )|^{2}}{V(\mathbf{r-r}%
^{\prime })}] \\
&&-Tr[\ln (-\mathbf{G}^{-1})].
\end{eqnarray*}

Using the saddle point condition, the Hartree-Fock self-energy and
superconducting gap are obtained as shown in Eq. (1).

\end{widetext}

\end{document}